\documentclass{aa} % use to 2 column version

\usepackage{latexsym}

\usepackage{natbib}
\usepackage{graphicx}
\usepackage{subfigure}
\usepackage{color}
\usepackage[dvipsnames]{xcolor}
\usepackage{txfonts}
\usepackage{bm} % bold greek letter

\graphicspath{{./}{figs-pdf/}}

\begin{document}

   \title{Miniature loops in the solar corona}

   %\subtitle{}

   \author{K.~Barczynski,
          \inst{1}
          H.~Peter
          \inst{1}
          \and S.L.Savage
           \inst{2}
          }

   \institute{\inst{1} Max Planck Institute for Solar System Research, 
           Justus-von-Liebig-Weg 3, 37077 G\"ottingen, Germany (barczynski@mps.mpg.de)\\
          \inst{2} NASA Marshall Space Flight Center, ZP 13, Huntsville, AL 35812, USA}

   \date{Received ...; accepted...}

% \abstract{}{}{}{}{} 
% 5 {} token are mandatory
 
  \abstract
  {Magnetic loops filled with hot plasma are the main building blocks of the solar corona. Usually they have lengths of the order of the barometric scale height  in the corona that is 50\,Mm.}
  {Previously it has been suggested that miniature versions of hot loops exist. These would have lengths of only 1\,Mm barely protruding from the chromosphere and spanning across  just one granule in the photosphere. Such short loops are well established at transition region temperatures (0.1\,MK), and we investigate if such miniature loops also exist at coronal temperatures ($>$1\,MK).}
  {We used extreme UV imaging (EUV) observations from the High-resolution Coronal Imager (Hi-C) at an unprecedented spatial resolution of 0.3\arcsec\ to 0.4\arcsec. Together with  EUV imaging and magnetogram data from the Solar Dynamics Observatory (SDO) and X-Ray Telescope (XRT) data from Hinode we investigated the spatial, temporal and thermal evolution of small loop-like structures in the solar corona above a plage region close to an active region and compared this to a moss area within the active region.}
  {We find that the size, motion and temporal evolution of the loop-like features are consistent with photospheric motions, suggesting a close connection to the photospheric magnetic field. Aligned magnetograms show that one of their  endpoints is rooted at a magnetic concentration. Their thermal structure, as revealed together with the X-ray observations, shows significant differences to moss-like features.}
  {Considering different scenarios, these features are most  probably miniature versions of hot loops  rooted at magnetic concentrations at opposite sides of a granule in small emerging magnetic loops (or flux tubes).}


 \maketitle
%
%________________________________________________________________

\section{Introduction}\label{S:introduction}

The major building blocks of the solar corona are loops. Observations of these structures have existed since the 1940s~\citep{1991plsc.book.....B}, with the key information acquired  through extreme UV (EUV) and X-ray observations. Coronal loops cover a wide range of temperatures and lengths;  from small transition region loops at 0.1\,MK being only a few Mm long \citep{2001A&A...374.1108P,2014Sci...346E.315H} to loops hotter than 10 MK and/or longer than 100\,Mm~\citep{2010LRSP....7....5R}. A typical active region loop would have a temperature of approximately 3\,MK and a length above 10\,Mm \citep{2010LRSP....7....5R}. Naturally, the question appears regarding the possible minimum length of a hot ($>$1\,MK) coronal loop. Magnetic field lines originating from very small bipoles might not reach above the height of the average chromosphere, which is, according to semi-empirical models, at some 2\,Mm.
Assuming a semi-circular geometry of the field line, this would correspond to a footpoint distance of 4\,Mm in the photosphere. However, the solar atmosphere is in a dynamic state, therefore one might expect loops even shorter than that.

The presence  small  bipolar magnetic structures carried upward either by granular convection or magnetic buoyancy was proposed by~\citet{1996ApJ...460.1019L} in the context of horizontal inter-network magnetic fields. Using spectro-polarimetry \citet{2007ApJ...666L.137C} showed that such low-lying magnetic loops can connect opposite magnetic polarities that are separated by only 2\arcsec\ in the photosphere. Such small magnetic bipoles can emerge, isolated in a transient fashion~\citep{2008A&A...481L..25I} with a mean lifetime of approximately 4 minutes~\citep{2009ASPC..415..132I}. In small-scale emergence processes the separation between the footpoints in the photosphere ranges from  0.5\,Mm
to 4\,Mm and is correlated with the lifetime of the emerging structure \citep{2009ApJ...700.1391M}. To reconstruct the three-dimensional structure of the magnetic field during the small-scale mergence, \citet{2010ApJ...713.1310I} investigated spectro-polarimetric data to invert the magnetic field vector as a function of height in the atmosphere.
They showed that the rising structures are indeed flux tubes with enhanced magnetic field reaching heights of 400\,km above optical depth unity at the surface (their Fig.\,8). Their diagnostics were limited to the photosphere, but it seems reasonable to assume that such a flux tube could rise  all the way to the top of the chromosphere if the emerging field was strong enough.

Indirect evidence for the existence of small cool transition region loops was first suggested by 
\citet{1983ApJ...275..367F}  based on spectroscopic data. In particular, he argued that part of the transition region emission originates  in unresolved fine structures, which would be cool loops not connected to the corona above. With the help of  spectral maps and spectroscopic investigations \citet{2000ApJ...535L..63W} and \citet{2000A&A...360..761P} investigated the properties of such transition region loops. Because of instrumental limitations, these could not be imaged directly in a clear fashion until the IRIS spectrograph and imager \citep{2014SoPh..289.2733D} became available. Using slit-jaw images, 
\citet{2014Sci...346E.315H} could follow the evolution of such cool loops with lengths of only a few Mm and lifetimes of a few minutes. Appearing in the quiet Sun network, considering their length, such loops would only span across one granule, probably connecting opposite magnetic polarities in the inter-granular lanes. 
%
%(2)
Such small cool transition region loops have also been investigated in one-dimensional models. \citet{2012A&A...537A.150S} showed that loops with lengths of some 1\,Mm to 15\,Mm could in principle explain the increase of the emission measure towards lower temperatures below $10^5$\,K. In their models the quasi-static loops always remained well below 1\,MK.

With the existence of small cool loops now being firmly established, 
the question is; can such small loops also reach higher  coronal temperatures?
One conceptual argument against this would be that a short magnetic fieldline with a length of
only one or a few Mm would still be covered by the chromosphere. Because of the high density there, it might be unlikely in terms of energy requirements to heat a significant amount of material to coronal temperatures.
However, recent spectroscopic observations have shown evidence of plasma in the dense transition from the photosphere to the chromosphere at approximately the temperature minimum perhaps being heated to almost 100\,000\,K \citep{2014Sci...346C.315P}. Therefore, it might well be that such structures are further heated to coronal temperatures.

Small  elongated structures with a footpoint distance of only 1 Mm reaching more than 1\,MK have been reported by~\citet{2013A&A...556A.104P}. Using data obtained from the High Resolution Coronal Imager \citep[Hi-C;][]{2013Natur.493..501C}, they found these structures to have a width of less than 200 km. These observations were only possible because of the high spatial resolution of the suborbital rocket experiment Hi-C. Its resolution is approximately five times higher than that of the current
workhorse of coronal imaging studies, the Atmospheric Imaging Assembly \citep[AIA;][]{2012SoPh..275...17L}, which
has a spatial resolution of approximately 1.4\arcsec\ corresponding to 1\,Mm and would not show those extremely small features.

 \begin{figure}
   \centering
  \includegraphics[width=8.8 cm]{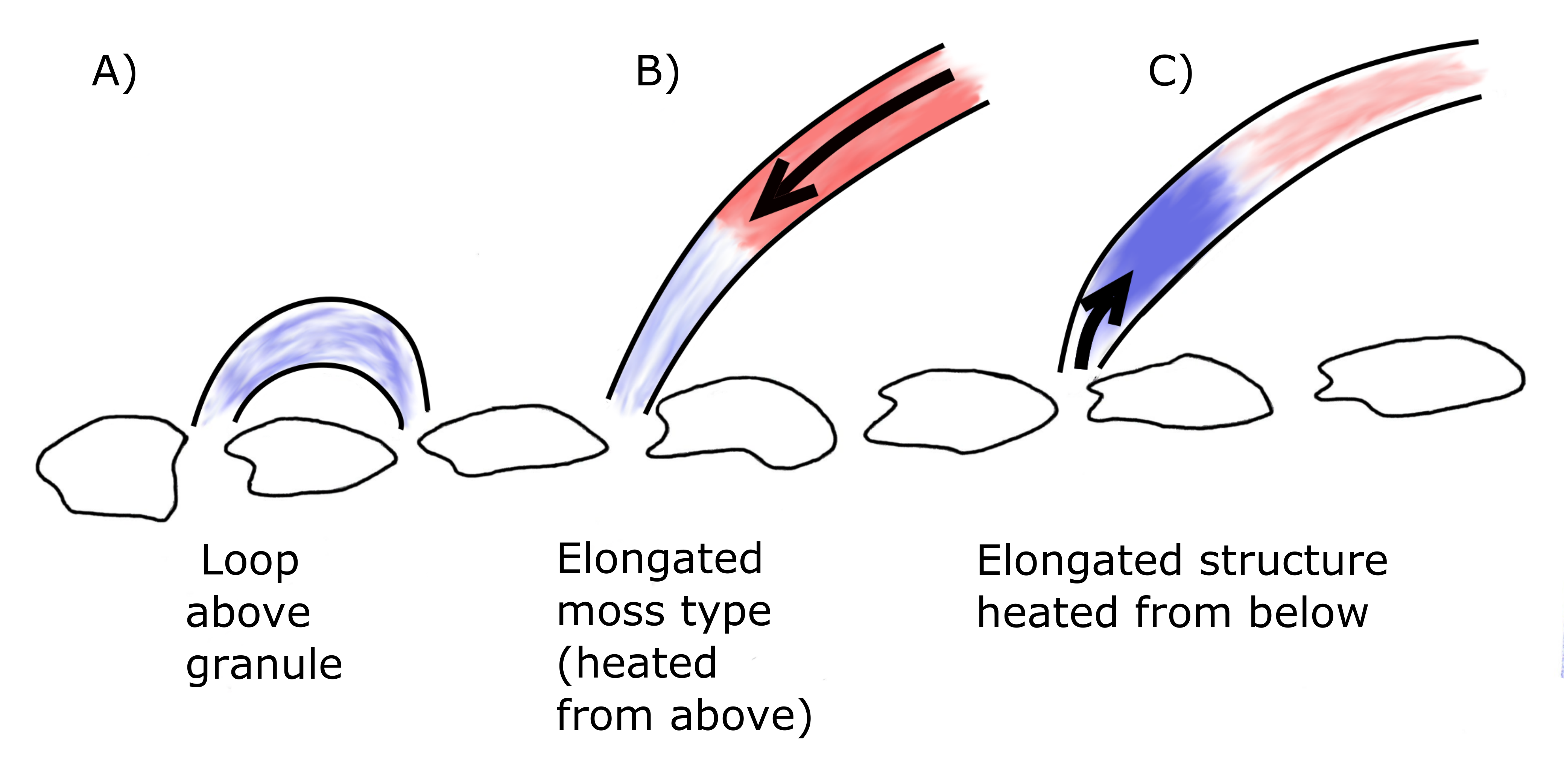}
      \caption{
Three scenarios for miniature loop-like structures seen between 1\,MK and 2\,MK.
(a) Small loops spanning across one granule, (b) short elongated structures at the footpoint of a hot loop, e.g.  moss, and (c) footpoint segments of a dilute hot loop heated from below (see Sect.\ref{S:introduction}).
The black solid lines  indicate granules.
The blue and red colours in the loop features denote warm (1\,MK to 2\,MK) and hot ($>$5\,MK) plasma.
The arrows represent the direction of the energy flux powering the bright feature, that is, in panel (b) the heat flux from the corona down to the lower atmosphere and in panel (c) the Poynting flux into the upper atmosphere.}
    \label{fig_0}
   \end{figure}

The small-scale coronal structures
have been interpreted as miniature loops by \citet{2013A&A...556A.104P}, spanning one granule (Fig.\,\ref{fig_0}a) and connecting two opposite magnetic field polarities in the inter-granular lanes. This would provide a natural explanation for their length, which is comparable to a granule, and be consistent with the emerging bipoles discussed above. Also, the motion of such miniature loops should be of the order of the photospheric horizontal motions which is typically of the order of 1\,km\,s$^{-1}$ \citep[e.g.][]{1975SoPh...40...53D}.
However, \citet{2013A&A...556A.104P} could not exclude an alternative scenario in which these elongated structures would be the moss-type emission (Fig.\,\ref{fig_0}b). Moss emission is characterised by dynamic arcsecond-scale features at the footpoint region of a hot loop typically reaching more than 5\,MK and being visible in X-rays \citep{1999ApJ...519L..97B}. In some sense the 1\,MK to 2\,MK emission near the footpoints represents the transition region of that hot loop. In the HiC data set \citet{2013ApJ...770L...1T} and~\citet{2014ApJ...789..105M} analysed moss structures but in the more active part of the HiC field-of-view, and not in the plage region where  \citet{2013A&A...556A.104P} found the small elongated structures.
There would also be a third option, where a longer structure is heated from below, filling only the lower part of the loop with 1\,MK to 2\,MK plasma near the footpoint with a dilute (basically invisible) hot part above (Fig.\,\ref{fig_0}c).

The small structures we  report here are quite different from coronal bright points, originally observed by \cite{1974ApJ...189L..93G} in X-rays. Those are much bigger with an overall average
size of approximately 30\arcsec{} and a bright core of approximately 5\arcsec{} to 10\arcsec
. They have much longer lifetimes of several hours and are typically associated with a bi-polar magnetic feature at the surface. However, there is a (magnetic) substructure in these features \citep{2001SoPh..201..305B}, and it has been suggested that a bright point might consist of small loops with widths of only approximately 1\arcsec\ to 2\arcsec\  and temperatures of approximately 1.6\,MK \citep{2008A&A...491..561D}. Nevertheless, it does not seem likely that the tiny loop-like features directly observed by \citet{2013A&A...556A.104P} and investigated here are related to the proposed elementary structures of a coronal bright point. This is supported by the magnetic structure, which is clearly bi-polar for a bright point, but mostly unipolar (perhaps with small-scale non-resolved opposite polarities) for the plage-type region hosting the very small loop-like features.

Our aim is to understand the nature of the tiny elongated structures identified by \citet{2013A&A...556A.104P} and to distinguish between the three scenarios outlined above and sketched in Fig.\,\ref{fig_0}. In particular, we investigate the morphology and evolution (Sect.\,\ref{S:properties}), the underlying magnetic field (Sect.\,\ref{S:magnetic.field}), and the thermal properties (Sect.\,\ref{S:thermal.structure}) of these features.

%__________________________________________________________________

\section{Observations}\label{S:obs}

In this study we concentrated on data acquired during a sub-orbital sounding rocket flight of the High-resolution Coronal Imager~\citep[Hi-C;][]{2013Natur.493..501C, 2014SoPh..289.4393K}. It was launched on 11th July, 2012, and acquired data for approximately
5 minutes of the active region AR11519-21 and its surroundings. For our analysis, we complement this (Table\,\ref{table:table1}) with data from the X-Ray Telescope onboard Hinode \citep[XRT;][]{2007SoPh..243...63G} and the Solar Dynamics Observatory (SDO); in particular the extreme UV images from the Atmospheric Imaging Assembly \citep[AIA;][]{2012SoPh..275...17L} and magnetograms from the Helioseismic and Magnetic Imager \citep[HMI;][]{2012SoPh..275..207S}.

%Hi-C part
The extreme UV imager Hi-C provides data of the solar corona in a 5\,\AA\ wide wavelength band around 193\,\AA\ dominated by emission from \ion{Fe}{xi} originating at approximately 1.5\,MK.  In the context of coronal imaging, HiC has an unprecedented plate scale of 0.1\arcsec\,pixel$^{-1}$, which corresponds to 73 km pixel$^{-1}$ on the Sun.  The image resolution estimated by~\citet{2014ApJ...787L..10W}
 is 0.3\arcsec\ to 0.4\arcsec. The 4k$\times$4k full-frame images were recorded with an exposure time of 2\,s at a cadence of 5.5\,s.  The unprecedented spatial and temporal resolution achieved by the Hi-C allows us to carry out the detailed analysis of the small-scale arcsec-size structures discussed in the introduction that were not resolved by previous instruments. We have studied 25 full-frame images (level-1.5) of the active region obtained between 18:53:11 UT and 18:55:30 UT on 11th July, 2012. All times in this paper are in seconds from this first frame.  We do not analyse the earlier 11 images (before 18:53:11 UT), because at this point the focus was not yet locked. The images after 18:55:30 contain only a partial read out of the detector (1k$\times$1k) acquired at a higher cadence, which unfortunately did not cover the region we are interested in here. The Hi-C images are aligned to sub-pixel level accuracy using a cross-correlation technique to obtain a jitter-free data set.

\begin{table}
\caption{Imaging data used in this study.}             
\label{table:table1}      
\centering                          
\begin{tabular}{l r c r@{}l c}        
\hline\hline                 
     &  band~~    & contribution & \multicolumn{2}{@{}l@{}}{$T_{\rm{peak}}$ [MK]\tablefootmark{(c)} }\\    \hline                      
Hi-C &  193\,\AA  & Fe XII   & 1&.5 \\ % http://solarscience.msfc.nasa.gov/papers/Cirtain/2013Cirtain_etal.pdf emission peak at 1.5MK for 193A channel
\hline  
     & 1600\,\AA & continuum & $<$0&.01 \\
     &  131\,\AA & Fe VIII   & 0&.4 \\
     &  171\,\AA & Fe IX     & 0&.7 \\
AIA\tablefootmark{(a)}
     &  193\,\AA & Fe XII    & 1&.5 \\ %http://www.aanda.org/articles/aa/pdf/2012/04/aa18144-11.pdf emission from Fe XXIV
     &  211\,\AA & Fe XIV    & 2&.0 \\
     &  335\,\AA & Fe XVI    & 2&.8 \\
     &   94\,\AA & Fe XVIII  & 7&.1 \\
\hline   
XRT\tablefootmark{(b)}
     &  Ti-Poly & free-free emission & ~~~~~9&\\
\hline                                 
\end{tabular}
\tablefoot{%
\tablefoottext{a}{All listed AIA channels are used to co-align the Hi-C observations with the HMI magnetograms, and all but the 1600\,\AA\ channel are employed in the DEM analysis (Sect.\,\ref{S:DEM}).}
\tablefoottext{b}{The XRT observations are used for thermal diagnostics.}
\tablefoottext{c}{For the HiC and AIA bands, $T_{\rm{peak}}$ is the temperature of peak ion fraction of the main contributing ion in the respective band according to \citet{2010A&A...521A..21O}. Most AIA bands have significant contributions from other temperatures as well. The value for XRT is taken from \cite{2007SoPh..243...63G}.}
}
\end{table}

\begin{figure*}[t]
   \centering
   \includegraphics[width=18.0 cm]{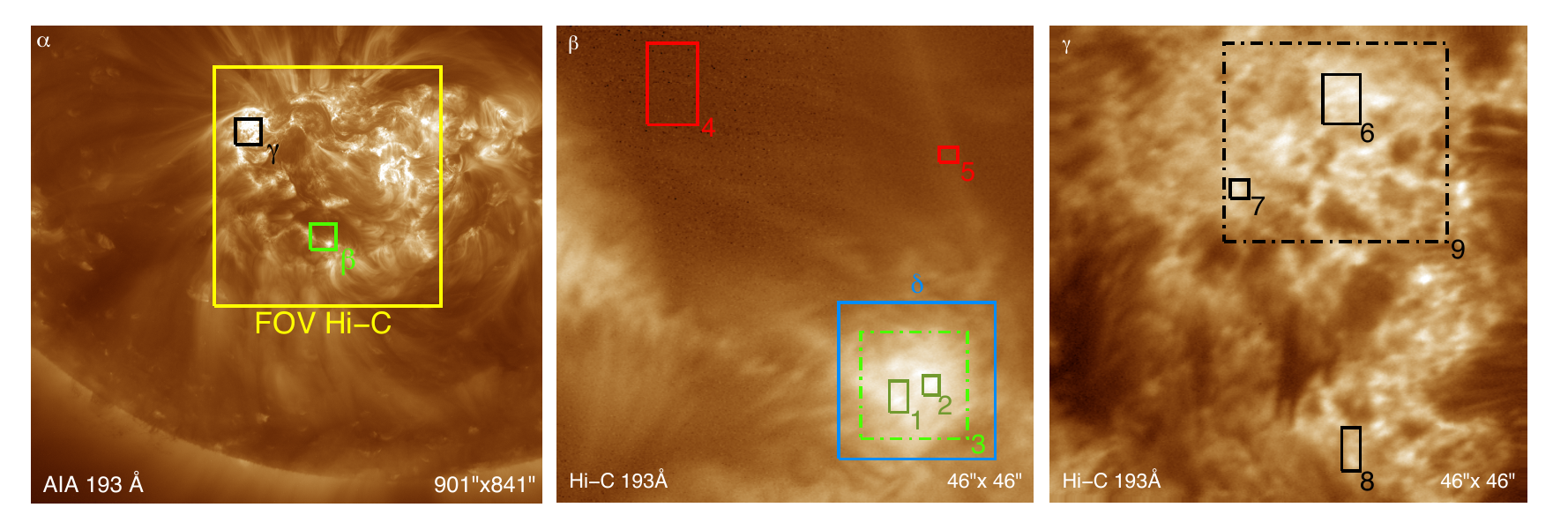}
      \caption{Active region context and regions of interest. All images are taken in the 193\,\AA\ channel showing emission around 1.5\,MK (see Table\,\ref{table:table1}). Panel \textbf{($\bm{\alpha}$)} shows part of the solar disk seen in AIA during the Hi-C rocket flight. The full field-of-view of Hi-C is indicated by the yellow box. The green and black boxes $\beta$ and $\gamma$ indicate  the plage area and the moss region displayed in panels $\beta $ and $\gamma$, respectively. Panel \textbf{($\bm{\beta}$)} displays a zoom into the plage region south of the active region (46"x46"). The boxes 1 to 5 indicates features used for the thermal study in Sect.\,\ref{S:DEM}. The box $\delta$ highlights the field-of view shown in Fig.\,\ref{fig_b}. Panel \textbf{($\bm{\gamma}$)} shows the zoom into a moss region. As in panel $\beta$, the numbered boxes show the features for the thermal study.} 
              \label{fig_a}%
    \end{figure*}

%AIA part
For context, for thermal studies (Sect.\,\ref{S:DEM}), and the alignment with the magnetic field data (Appendix\,\ref{S:alignment}), we used data from various channels of AIA (cf. Table\,\ref{table:table1}).
This provided full-disk images at a plate scale of 0.6\arcsec\,pixel$^{-1}$ at a cadence of 12\,s or more. In Fig.\,\ref{fig_a} we show the context of the active region observed by Hi-C as well as our regions of interest. In this paper we concentrate on two small plage-type regions that are shown in Fig.\,\ref{fig_a}$\beta$  and $\gamma$ along with areas marked for a more detailed study.

%HMI part

To investigate the relationship between the small loop-like features and the surface magnetic field, we use magnetogram data from HMI. Provided with a plate scale of 0.5\arcsec\,pixel$^{-1}$ and a temporal cadence of 45\,s, HMI provides only a few snapshots during the 140\,s of the full-frame Hi-C data used here. Thus, we concentrate on the line-of-sight magnetogram taken at 18:54:53 UT in the middle of the Hi-C full-frame coverage. The spatial alignment between HMI and Hi-C was achieved through a sequence of AIA channels using a cross-correlation procedure (see Appendix\,\ref{S:alignment}). We estimate the accuracy of this alignment to be better than 0.2\arcsec\ (Table\,\ref{table:2}), that is, slightly more than half a pixel of HMI.
While magnetic field information at higher spatial resolution and magnetic sensitivity than that available with HMI would have been desirable, unfortunately, such data were not available for our region of interest, either from the ground or from the space-based Hinode observatory.

%XRT-part
To study the possible presence of very hot plasma in relation to the small loop-like features, we used data from XRT that provides a spatial scale of 1\arcsec\, per pixel. During the Hi-C flight XRT, data were taken with the Ti-poly filter showing plasma in a broad temperature with a peak at approximately 9\,MK. From the peak down to a temperature of 2\,MK, the response of Ti-poly drops by a factor of approximately 15 \citep[][their Fig.\,7]{2007SoPh..243...63G}. Unfortunately, the plage area of interest in our study is only partially covered by the XRT field-of-view.  Still, the region connecting the plage to the main part of the active region is covered, which makes the XRT data very valuable for our study (see Sect. 5.2).

\section{Properties of small loop-like features}\label{S:properties}

Small loop-like features have been found in Hi-C observations by \citet{2013A&A...556A.104P}. Based on a single image they   found that these have a length of approximately 1.5 Mm and a width below 200 km, and they suggest an interpretation of them as miniature versions of coronal loops. The main purpose of our study is to investigate the spatial and temporal evolution and the thermal structure of these features and to relate them to the underlying magnetic field structure. We first discuss length, width and relative orientation with respect to the E-W direction of the Sun and horizontal motions of the loop-like structures as a function of time.

\subsection{Identification of small loop-like features}\label{S:identification}

\begin{figure*}
  \sidecaption
 \includegraphics[width=12.0 cm]{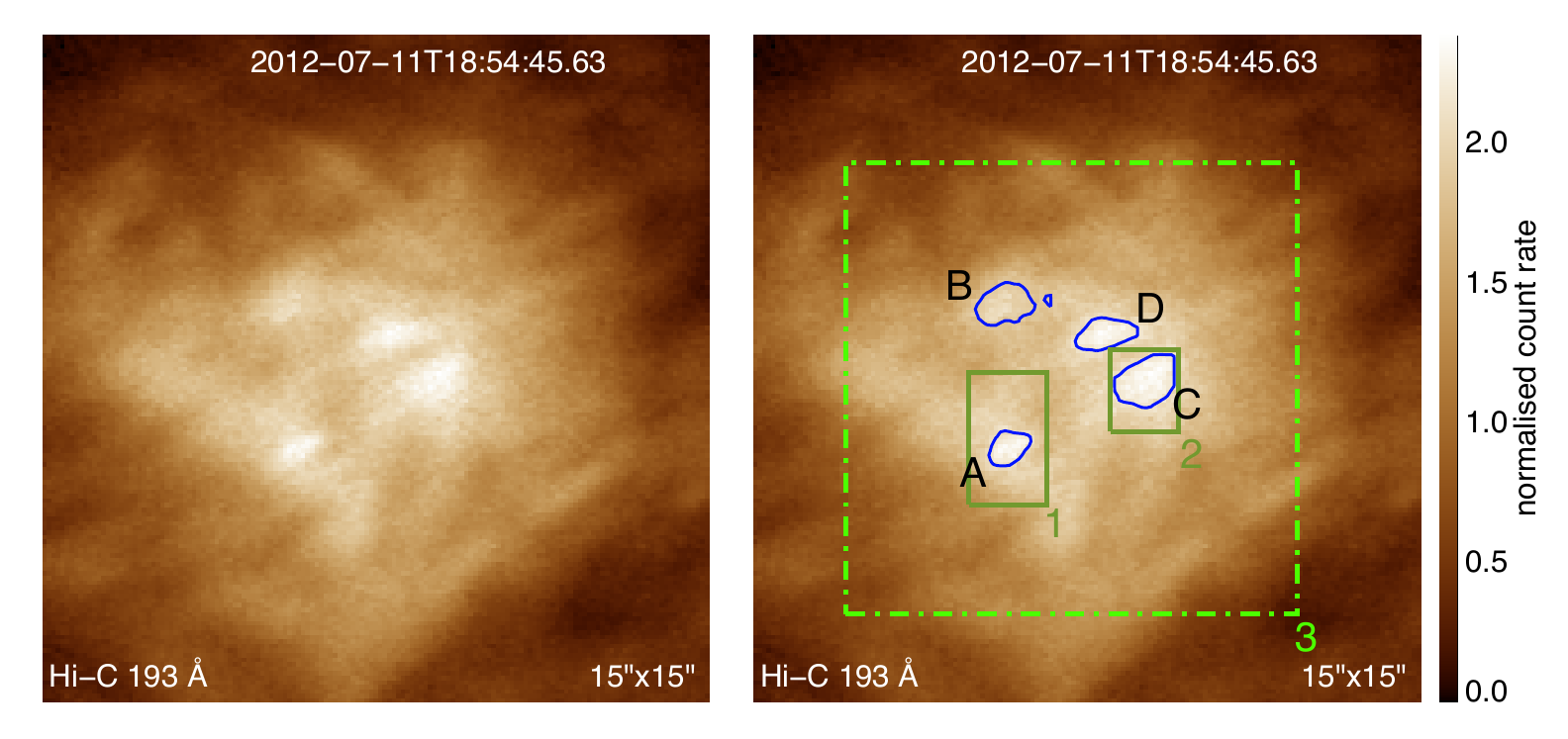} 
      \caption{Zoom of the plage area seen by Hi-C (see box $\delta$ in Fig\,\ref{fig_a}$\beta$). We show the same image with and without annotation. The contours A to D show the miniature loop-like features identified with the procedure outlined in Sect.\,\ref{S:identification}. The boxes 1 to 3 show the features for the thermal analysis in Sect.\,\ref{S:DEM} (same boxes as in  Fig.\,\ref{fig_a}$\beta$). }
              \label{fig_b}%
    \end{figure*}

Identification of the loop-like features is via a combination of manual and automated procedures. Firstly, on each of the 25 images, we identified the 15"x15" subregion hosting the features of interest (Fig.\,\ref{fig_b}). For each image, the intensity in that subregion of 15"x15" was normalised by the respective median value of intensity from that region. Then, the contrast of the subregion is enhanced by employing a median filter with a kernel size of 3\arcsec.

We approximately identified four features by eye, marked A to D in Fig.\,\ref{fig_b}. For each structure we defined  a rectangular subfield only just covering this feature and its immediate surroundings (e.g.\ box 1 in  Fig.\,\ref{fig_b} for feature A). In each of the small boxes, we calculated the maximum of the normalised and contrast-enhanced brightness, $I_{max}$, during the time series. We then defined the small loop-like structures as the feature enclosed by the contour line at  a level of 90\% of $I_{max}$ (cf. contours A to D in Fig.\,\ref{fig_b}).        

 To describe the properties of the features we fitted an ellipse%
\footnote{%
For ellipse fitting we use the function mpfitellipse from the IDL library of Craig Markwardt
(\url{https://www.physics.wisc.edu/~craigm/idl/}).
}
to the respective contour lines at a level of  90\% of $I_{max}$. This is motivated by the observation that the contour lines have an approximately ellipsoidal shape.
The properties of the fitted ellipse in each time step was then used to characterise the loop-like features: the position of the ellipse represents the position of the loop feature; for its length and width we use the major and minor axis of the ellipse, the angle is defined as the angle of the major axis with solar-x, that is, the E-W direction and the brightness of each feature was calculated by the mean brightness within the ellipse. This was done independently for each of the exposures of the time series of full-frame Hi-C data so that we could also investigate the temporal evolution, at least over the section covered by more than two minutes of the full-frame data.
To study the brightness variability we used images without contrast enhancement by the median filtering, of course.

The spatial extent of the loop-like features, as characterised by the ellipse contours at 90\% intensity level, is a measure for the full width at half maximum of the features. While the plage region shows a peak intensity of a factor of approximately 10 higher than the surrounding quiet areas, the individual loop-like features have an intensity contrast of approximately 20\% above the plage region in the immediate vicinity of them \citep[see Fig.\,4 of][]{2013A&A...556A.104P}. Therefore, the 90\% level of the peak intensity (without background subtraction) represents the full width at half maximum. The values derived for the widths and lengths of the loop-like features can be lower limits only, because the background emission from the plage region might hide a larger (low intensity) extension of the features. However, this argument applies to any measurement of, for example,\ loop widths in the corona.

 \subsection{Properties of small loop-like features}\label{S:prop.detail}

\begin{figure}
   \centering
   \includegraphics[width=8.8 cm]{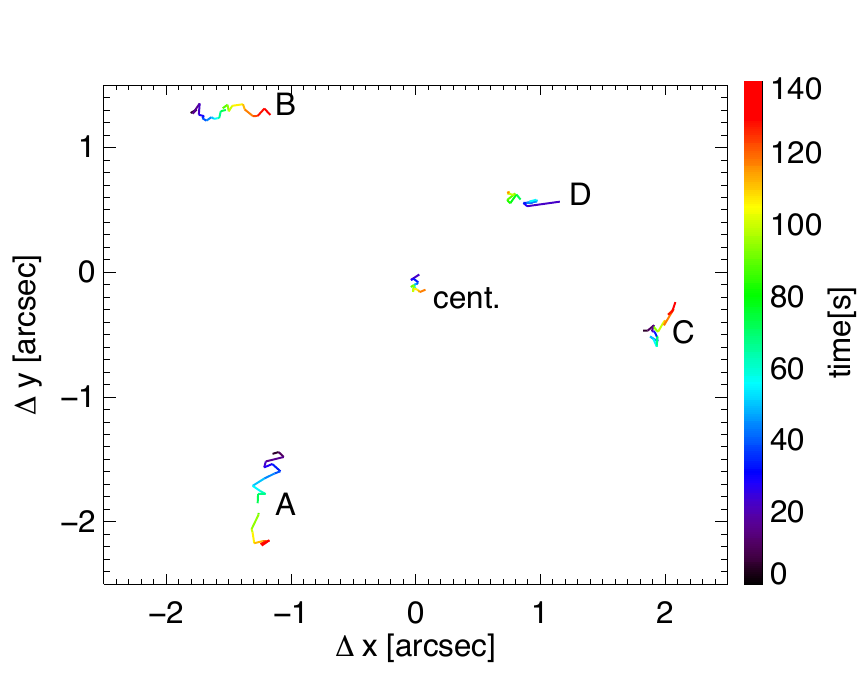}
      \caption{Horizontal motion of small loop-like features. The trajectories
A to D show the centre of the respective features identified in Fig.\,\ref{fig_b}
 as a function of time (colour-coded) over 140s (after 18:54:11 UT). The
error estimate for  the position from the ellipse fitting procedure is approximately
0.5 pixels or 0.05" (half a tick mark). For comparison, the trajectory labelled
'cent.' shows the mean position of the four loop-like features A to D. See
Sect.\ref{S:identification}.
 } 
              \label{fig_d}%
    \end{figure}

The position of the features slowly changes in space, typically over slightly less than 1\arcsec\ over the 140s covered by the full-frame images. The positions of the four features  and their evolution in time is shown in Fig.\,\ref{fig_d}. To check that there are no systematic residual image motions, we check the mean position (equal weighting) of the four intensity features that is found in the middle of Fig.\,\ref{fig_d}. This is almost stable with a motion of ${<}0.2$\arcsec\ in 140s ~($ \cong1.0$ km s$^{-1}$). This reflects the good temporal alignment
of the Hi-C data. The  trajectories of the four features (A to D) cover approximately 1\arcsec\ during the observation in a more or less linear fashion. The motion of ${\approx}
1"/140$s corresponds to a speed of 5\,km\,s$^{-1}$. That all four features move in random directions (as far as can be judged from four data points) underlines the idea that
no systematic effects cause this result.
In general, these structures exhibit random motions,
which are consistent with the motion of the small-scale magnetic loops~arching over granules discussed in the introduction.
In particular, this horizontal velocity is consistent with observations of flux emergence on granular scales.
As shown by \citet{2009ApJ...700.1391M}, in those cases, footpoints can be separated by approximately 2\arcsec\ within four minutes (see their Fig.\,2), which would be approximately 6\,km/s.
Thus, the average velocity of the features we see is approximately comparable to average horizontal photospheric motions under small-scale flux-emergence conditions.

\begin{figure*}
   \centering
  \includegraphics[width=18cm]{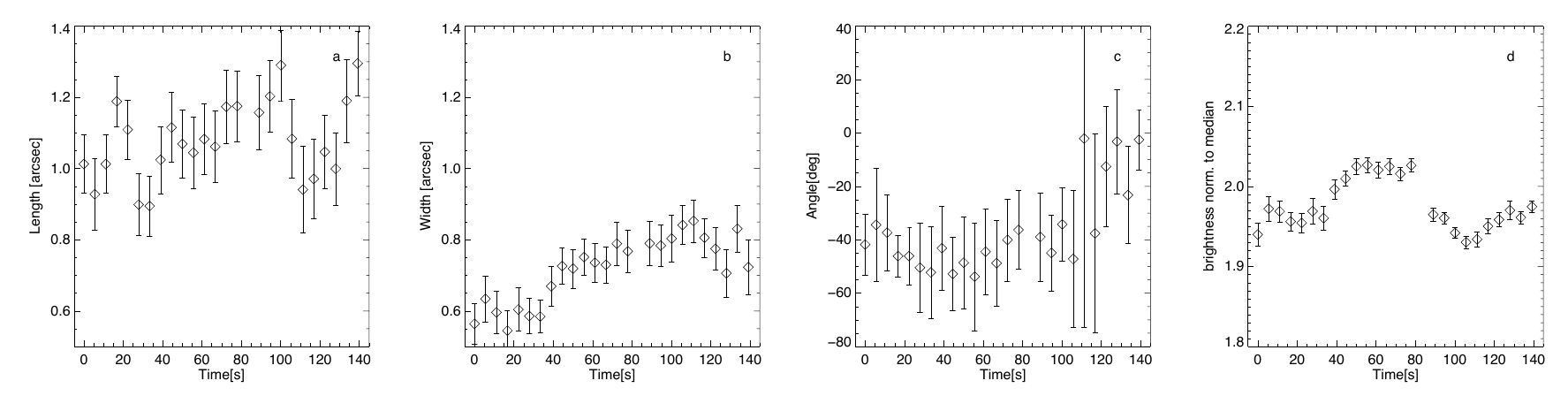}
      \caption{Temporal variation of the properties of one loop-like feature. This plot is for feature B as marked in Fig.\,\ref{fig_b}. The properties of the other three features are shown in Figs.\,\ref{str_a} to \ref{str_d}. These data are based on the ellipse fits of the features: (a) the major axis representing the length, (b) the minor axis the width, (c) the angle of the major semi-axis with the E-W direction for the angle, and (d) the intensity integrated over the ellipse for the structure brightness. The time is with respect to the first full-frame Hi-C image at 18:53:11 UT.
See Sect.\ref{S:identification}. }
         \label{fig_e}
   \end{figure*}

In general, while the structures move, they also (slightly) change their length and width. This is illustrated in  Fig.\,\ref{fig_e}a,b for feature B  (for the other
features see appendix\,\ref{S:app.properties}, Figs.\,\ref{str_a} to \ref{str_d}). The typical length and width of the loop-like structures are 1\,Mm and 0.5\,Mm, respectively, with a typical aspect ratio of approximately two. Strictly speaking, the footpoint distance of approximately 1\,Mm is only representative of the loop length if it is a low-lying (loop) structure (which we implicitly assume here). If it were a semi-circular loop, its length would be $\pi/2$ times the footpoint distance, that is, some 50\% longer.

 Besides the horizontal motion of the features as a whole (Fig.\,\ref{fig_d}), we also follow the motion of their endpoints
(viz. footpoints)  relative to their  center. This is done
by measuring the angle of the fitted ellipse with respect to the E-W direction.  As an example, Fig.\,\ref{fig_e}c
shows the change of this angle during the Hi-C observation for  feature B (again, other features in Figs.\,\ref{str_a} to \ref{str_d}).
Because of the considerable uncertainties in determining the angle (because the ratio of length to width is not big), the variation of the angle is almost within the error bars. Nevertheless, in three of the four cases we see a gradual rotation
of the loop feature, with a difference between the maximum and minimum angles of approximately $\alpha{\approx}40^\circ$ (expect the case in Fig.\,\ref{str_a}c). While the question remains as to the significance of this value, we can at least estimate this as an upper limit.
Considering the length of approximately 1\,Mm, this angle corresponds to a rotational component of the motion of approximately $\sin(40^\circ)\times1\,{\rm{Mm}}\approx0.6\,{\rm{Mm}}$. Thus the upper limit of the speed of the footpoints during the 140s of the Hi-C full-frame data is approximately 4\,km\,s$^{-1}$.
This is approximately consistent with the velocity of magnetic concentrations in the intergranular lanes derived from magnetic bright points \citep[][their Fig.\,2]{2014A&A...563A.101J}     and supports the interpretation that the footpoints are rooted in flux tubes in intergranular lanes.

Finally, we verified the variability of the brightness in time. For feature B this is shown in Fig.\,\ref{fig_e}d. There is a variation of approximately 5$\%$ (for other features, even up to 15$\%$) of the brightness over approximately 100s. This variation of the brightness is significant in two respects. Firstly, this variability is approximately four times larger than the average error in brightness (taking into account the photon noise and the error in the ellipse fitting shown as bars in the plots). Secondly, this variation of 5\% to 15\% is also significant when considering that the contrast of the loop-like features above the background of the plage is only some 20\% (see end of Sect.\,\ref{S:identification}). This implies, that the change of the emission from the loop-like feature alone would change by 50\% to almost 100\% during the observation time. Still, the loop-like features are visible just above the background throughout the full Hi-C observation sequence of approximately two minutes.
Naturally we can not draw a final conclusion on the lifetime of the loop-like features because the length of the time series of the full-frame Hi-C images is limited, but at least we can say that the lightcurve is consistent with a feature lifetime of two minutes or more. For the other features, the situation is less clear (Figs.\,\ref{str_a} to \ref{str_d}). One feature shows almost no intensity variation, while the other two are caught while the intensity decreases during the observation time. We can thus conclude that the features probably have a lifetime of a few minutes or more.

Summarising these findings, the observed features show motions and lifetimes consistent with being anchored in magnetic concentrations in intergranular lanes. This would be consistent with the loop-like features being indeed miniature hot loops spanning one granule, and them being the result of rising magnetic flux tubes of the type reported by  \citet{2010ApJ...713.1310I}, for example. Furthermore their properties in terms of length, width and lifetime are comparable to the transition region loops at 0.1\,MK reported by \citet{2014Sci...346E.315H}, therefore it might be that the features we see here are a coronal temperature version of those transition region loops.

However, based on the above information, it could still be that we see only one footpoint of a longer loop, similar to the scenarios discussed in the introduction (cf. Fig\,\ref{fig_0}).
For this, we discuss their relation to the magnetic field (Sect.\,\ref{S:magnetic.field}) and the thermal structure (Sect.\,\ref{S:thermal.structure}).

%__________________________________________________________________

\section{Relation to photospheric field}\label{S:magnetic.field}

%-------------------------------------------------------------------------------

\begin{figure}
   \centering
   \includegraphics[width=8.8 cm]{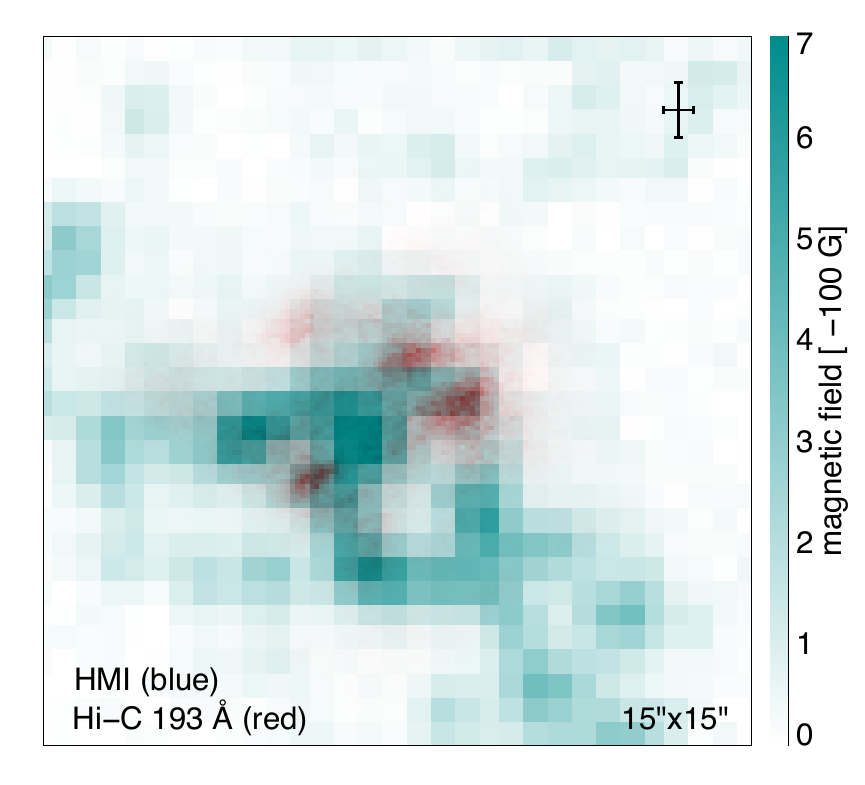}
      \caption{Aligned composite of photospheric magnetogram and coronal emission. In the same field-of-view as in Fig.\,\ref{fig_b}  we plot the magnetogram on a blue scale and overlay the coronal image from Hi-C in red. The magnetogram in this field-of-view contains only one single magnetic polarity. The Hi-C image is the same as in Fig.\,\ref{fig_b}, but on a non-linear colour scale to emphasise the loop-like features. See Sect.\,\ref{S:magnetic.field}. }
              \label{fig_f}%
    \end{figure}

For an investigation of how the loop-like features relate to the magnetic field, we carefully aligned the Hi-C images with the HMI magnetograms (for details of the alignment see Appendix\,\ref{S:alignment}). The accuracy of the alignment of the HMI magnetogram with the Hi-C image is within less than half a pixel of HMI, at approximately 0.2\arcsec.  The result of the alignment is displayed as a composite in Fig.\,\ref{fig_f} showing the magnetogram along with the Hi-C image of coronal plasma over-plotted in red.

Within the sensitivity and resolution
of HMI, only one polarity is seen in the magnetogram of this plage region, as noted already by \citet{2013A&A...556A.104P}
in their study of this same region. In the field-of-view in Fig.\,\ref{fig_f} not a single pixel of the HMI magnetogram shows an opposite polarity. However, this does not exclude the presence of small parasitic opposite polarities.  At higher spatial resolution and magnetic sensitivity, one might well see opposite polarities not visible to HMI. For example, data from the Imaging Magnetograph eXperiment \citep[IMaX;][]{2011SoPh..268...57M}
 on the Sunrise
balloon telescope \citep{2010ApJ...723L.127S},  at much higher resolution and sensitivity than HMI, clearly show small-scale magnetic patches of opposite polarity close to a largely unipolar region  
\citep[e.g.][their Fig.\,2]{2013SoPh..283..253W} that would not be detectable by HMI. Unfortunately, as discussed in Sect.\,\ref{S:obs},  no observations of higher resolution or sensitivity are available for this region of interest.

The comparison of the Hi-C image and the magnetogram in Fig.\,\ref{fig_f}
shows that typically at least part of the loop-like feature is located above a concentration of magnetic field. According to the scenarios in Fig.\,\ref{fig_0} this would support the theory that one (or the only) footpoint is rooted in the photospheric magnetic field concentration. Because of the arguments concerning weak, small and opposite-polarity patches, the missing detection of HMI of a possible second footpoint cannot be used as an argument against the small-scale loop scenario.
Therefore this discussion of the relation of the magnetic field to the emission seen in Hi-C must remain indecisive and we must rely on the analysis of the thermal structure at and around the plage region hosting the small loop-like features.

\section{Relation to thermal structure}\label{S:thermal.structure} 
   
There is a significant difference between a miniature coronal loop and the
moss emission that might also appear as an elongated structure (Fig.\,\ref{fig_0}a,b):
by its very nature in the vicinity of the moss structure there should be
some signature of the hot loop that is connected to the moss region, while
for an isolated miniature loop one would not expect that hot component of
the emission. To address this difference, we investigate the differential
emission measure (DEM; Sect.\ref{S:DEM}) and the X-ray emission (Sect.\,\ref{S:X.rays})
from the plage region hosting the small loop-like features and compare it
to a moss area.

\subsection{Differential Emission Measure (DEM)}\label{S:DEM}

%(4*)

    \begin{figure*}
   \centering
   \includegraphics[width=18.0 cm]{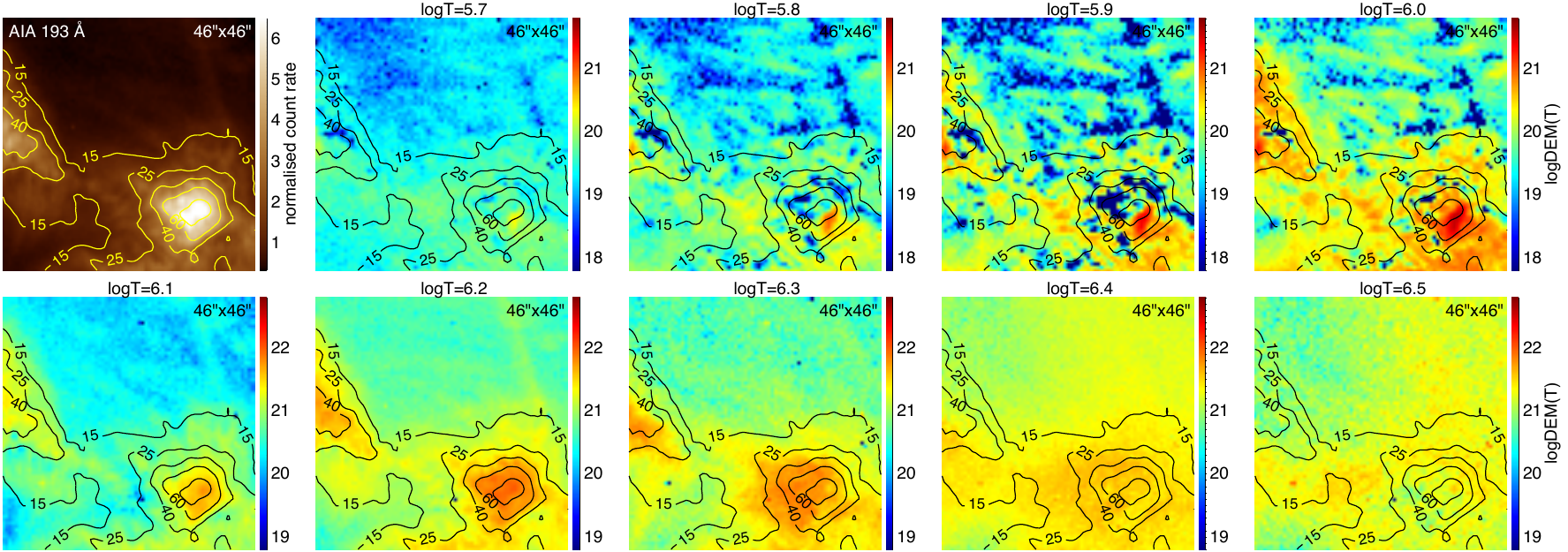}
      \caption{Coronal image and thermal structure. The top left panel shows the AIA image at 193\,\AA. The other panels   display spatial maps of the differential emission measure (DEM) at different temperatures $T$ labeled by ${\log}\,T$\,[K]. All panels show the same field-of view that is identical to Fig.\,\ref{fig_a}$\beta$.
  The plage region with the miniature loop-like structures is in the bottom-right part. For a better comparison between the panels we add contour lines of the brightness in the AIA 193\,\AA\ channel to all the panels.  The numbers with the levels denote the percentage of the peak 193\,\AA\ brightness in this field-of-view. All  DEM maps have the same dynamic range of $10^4$ although their lower (and upper) limits change to best represent the DEM structures.
See Sect.\,\ref{S:DEM}.} 
              \label{fig_g}%
    \end{figure*}

We perform a DEM analysis in the vicinity of the loop-like features and compare
this to a moss region  that has been analysed before 
\citep{2013ApJ...770L...1T} as well as to quiet regions without notable emission in the 193\,\AA\ band for a blind
test, all in the Hi-C field-of-view. The DEM provides some information on how the plasma is distributed
in temperature throughout the atmosphere (along the line-of-sight) and is
defined as
\begin{equation*} \label{eq1}
DEM=n^{2}_{e}\left( \frac{dT}{dh}\right)^{-1} ~.
\end{equation*} 
Here $n_e$ is the electron density, $T$ the temperature, and $h$ the height
along the line-of-sight. Being a function of $T$, the DEM is a measure
for how much plasma is present
at temperatures where the EUV (and X-ray) emission is originating.
Hi-C alone can not be used to calculate the DEM because it provides only
one wavelength band. In contrast, AIA includes a sufficient number of  bands
spanning the typical temperatures in the corona, albeit at a significantly
lower spatial resolution. So we use AIA imaging data (see Table\,\ref{table:table1}) to perform an inversion resulting in maps of the DEM over the covered temperature range.
This analysis is based on a set of near-simultaneous images obtained between 18:53:56 and 18:54:06, one in each channel. In our analysis we took into account photon noise and readout noise.
\footnote{%
To calculate the photon noise and the read noise of the SDO/AIA images we use the procedure aia\_bp\_estimate\_error available in SolarSoft
(\url{http://www.lmsal.com/solarsoft/}).
}%

There are numerous publicly available methods  to perform this inversion,
and we decided to use the regularised DEM inversion by 
\citet{2012A&A...539A.146H} for its robustness.

\begin{figure}
   \centering
   \includegraphics[width=8.8 cm]{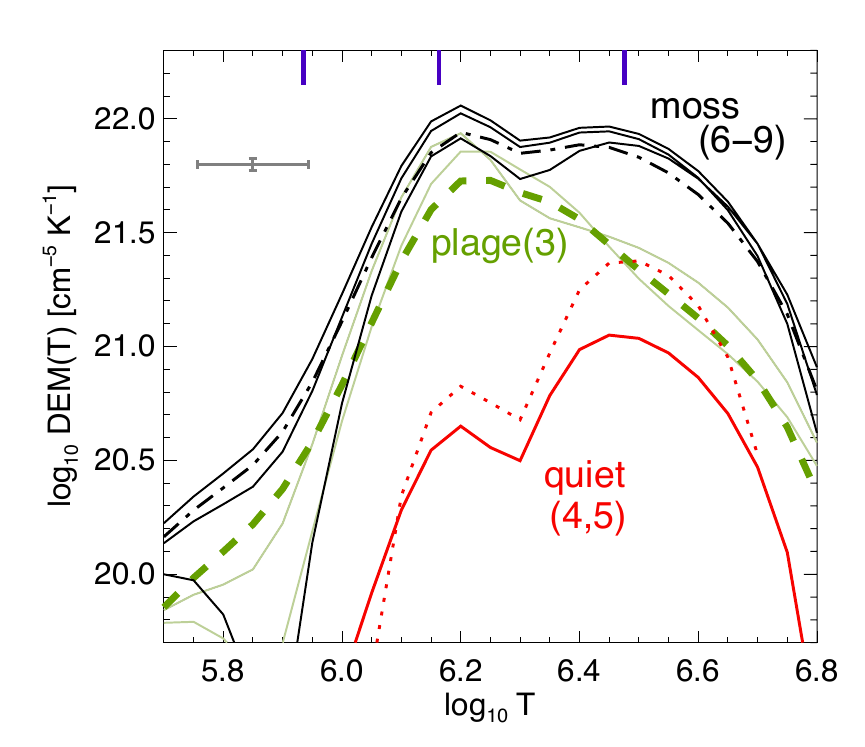}
      \caption{ Differential emission measure (DEM) distribution for different structures. The labelling of the lines corresponds to the regions marked in Fig.\,\ref{fig_a}
$\beta ,\gamma$. The green lines represent two individual small loop-like features (thin green; 1, 2) and an average covering the whole plage area hosting the small loop features (thick dashed green; 3). For comparison the red lines are the inversions for the quiet regions (4, 5). The black lines show the DEM in the moss areas (6-8), with the thick dashed line for the average over the larger moss region (9). The error bars (grey) present the average uncertainties for all curves in logT range from 6.0 to 6.7. The three blue markers at the top axis indicate the temperature of the peaks of the DEM components in coronal holes, quiet Sun, and active region according to \citet{2008ApJ...672..674L}. See Sect.\,\ref{S:DEM}.} 
              \label{fig_h}%
    \end{figure}

To obtain DEM curves for each region-of-interest (see boxes 1-9 in Fig.\,\ref{fig_a} $\beta ,\gamma$ ) we calculated the average intensity in each region-of-interest for each AIA channel.
The inversion procedure as described by \citet{2012A&A...539A.146H} uses these AIA count rates as input and provides the DEM as a function of temperature, including error estimates for the temperature and the DEM (shown as a cross in Fig.\,\ref{fig_h}).
One contribution to the errors in the DEM are the uncertainties in the count rates of the AIA channels (with counts typically ranging from 50\,DN/pixel/s in the weak channels (e.g. 94\,\AA) to a few thousand \,DN/pixel/s in the stronger channels (e.g. 193\,\AA).
However, the errors returned by the procedure also include uncertainties of the method and the width of the contribution functions of the AIA channels in temperature \citep[for details see][]{2012A&A...539A.146H}.

The maps of the DEM in the vicinity of the small loop-like structures (Fig.\,\ref{fig_g}) show a clear signature of an enhanced DEM at temperatures around 1.5\,MK $({\log}T{=}6.2$; here and in the following, all ${\log}T$ values refer to $T$ in units of K).
This simply reflects the enhanced emission we see in the Hi-C passband. In particular, the DEM maps do not show any significant signal at higher temperatures.

To emphasise this result from the spatial maps, we check the DEM curves as a function of $T$ for selected regions (Fig.\,\ref{fig_h}).
In the areas directly above two of the loops (regions 1 and 2 in Figs.\,\ref{fig_a}$\beta$ and \ref{fig_b}) as well as in a larger area encompassing all the small loop-like features in the plage region (region 3 in Figs.\,\ref{fig_a}$\beta$
and \ref{fig_b}) we see a clear peak at approximately ${\log}T{=}6.2$. The  drop to lower and higher temperatures is comparably sharp. In particular, the drop to higher temperatures distinguishes the loop-like features from the moss area (see below). As expected, the DEM from the plage region is significantly higher than in sample regions covering the quiet regions (see Fig.\,\ref{fig_h}). In particular, also region 5 in  Fig.\,\ref{fig_a}$\beta,$ that is located at a faint longer loop seemingly connecting the plage area with the main part of the active region to the North, does not show any enhanced DEM at higher temperatures, at least not higher than the DEM curves for the plage regions 1 to 3. This is why we labelled this region 5 as 'quiet', as well, in Fig.\,\ref{fig_h}.

To confirm that this technique would detect the presence of hot plasma in moss regions, we checked the moss region in the Hi-C field-of-view that has been investigated already by \citet{2013ApJ...770L...1T}. These are regions 6 to 9 in Fig.\,\ref{fig_a}$\gamma$. The DEM curves of these regions in Fig.\,\ref{fig_h} are distinctively different from the loop-like features in the plage. While the DEM of the moss regions is comparable around ${\log}T{=}6.2$, where the plage region peaks, only the moss regions show a clear enhancement of the DEM up to at least ${\log}T{=}6.6$.

Therefore we conclude that there is no hot plasma present in the vicinity of the loop-like features. At least not to the same extent as in moss regions. To further verify this, we also investigated the X-ray emission.

\subsection{X-Ray observations}\label{S:X.rays}

     \begin{figure*}
   \centering
   \includegraphics[width=18.0 cm]{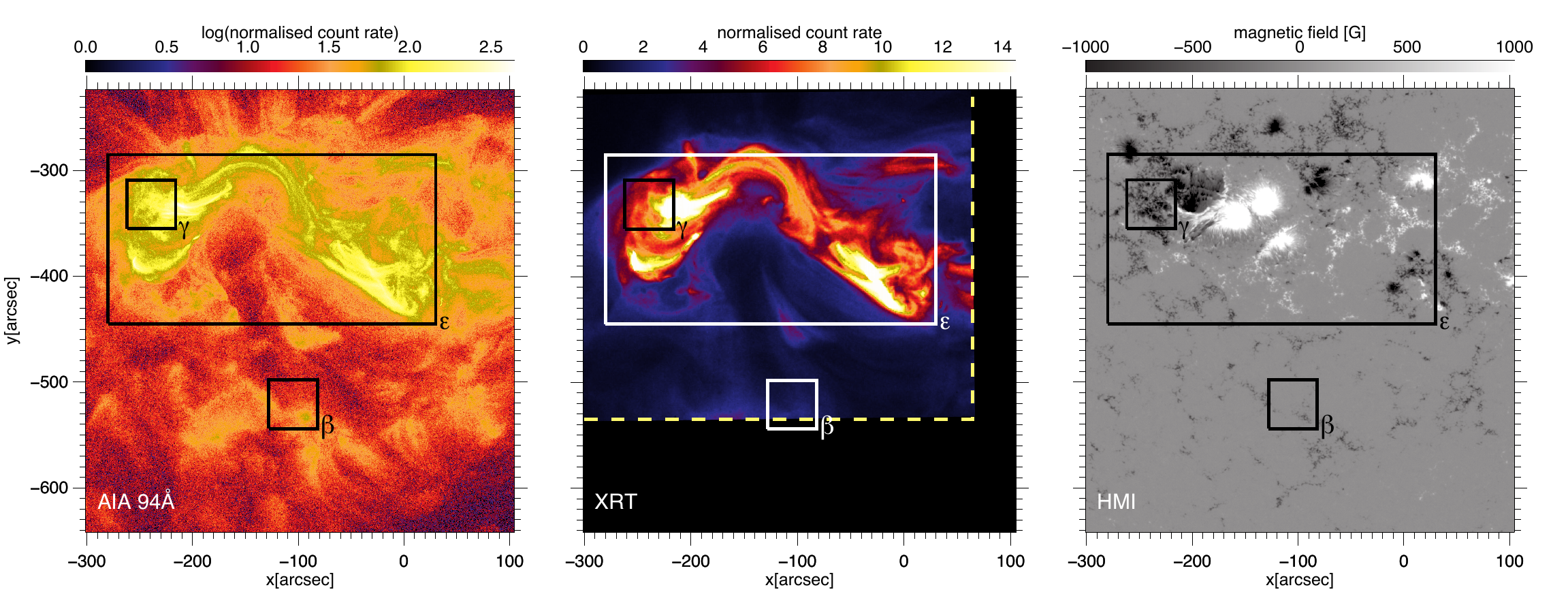}
      \caption{Emission from hot plasma and underlying magnetic field. The
left two panels show the emission in the AIA channel at 94\,\AA\ and the
Ti-poly channel of XRT on Hinode. The XRT image is taken during the Hi-C
flight (at  18:54:48
UT). The right panel shows the HMI magnetogram. All panels cover the full
field-of-view of the Hi-C data (cf. large rectangle in Fig.\,\ref{fig_a}$\alpha$).
  The box $\beta$ indicates the region with the plage area hosting the small loop-like
features as also displayed in Fig.\,\ref{fig_a}$\beta$.  The rectangle 
$\varepsilon$ outlines the area covering hot coronal loops and moss as investigated
by Testa et al. (2013). In their paper they display this region $\varepsilon$
rotated counter-clockwise by 90$^\circ$ The moss area displayed in Fig.\,\ref{fig_a}$\gamma$
and marked here also by $\gamma$ roughly corresponds to the moss region M3
of Testa et al. (2013), their Fig.\,2. The yellow dashed line in the XRT
image marks the edge of the XRT field-of-view. See Sect.\,\ref{S:X.rays}.}
              \label{fig_i}%
    \end{figure*}

Including X-ray observations can help to relax the limitation of AIA in terms of temperature coverage. While AIA alone can provide the DEM only up to approximately ${\log}T{=}6.6$ or 6.7 (${\approx}5$\,MK; cf.\ Fig.\,\ref{fig_h}), XRT observations cover a broad range of temperatures centered at almost 10\,MK (cf.\,Table\,\ref{table:table1}). In principle one could include X-ray observations in the DEM inversion (e.g.~\citet{2015ApJ...807..143C}. Here we choose the more straight forward way and  study the X-ray images directly. This provides information if a hot component of a loop rooted in the loop-like features is present that might not be revealed by AIA. The DEM, as derived from AIA in the plage region hosting the small loop-like features, drops steeply above ${\log}T{=}6.3$ (${\approx}2$\,MK). The response of the Ti-Poly filter of XRT drops  by a factor of (only) approximately   15 from its peak (at 9\,\,MK) to 2\,MK \citep[][]{2007SoPh..243...63G}. Therefore, XRT is well suited to test if there is a hot component of plasma that would go unnoticed by the DEM analysis of AIA. The hot emission from X-rays in the Hi-C field-of-view as seen by XRT on Hinode is displayed in Fig.\,\ref{fig_i}. For comparison we  show the AIA\,94\,\AA\ channel that reveals  comparably hot plasma (cf.\,Table\,\ref{table:table1}), hotter at least than 2\,MK and colder than 12.5\,MK. In this context we used the AIA\,94\,\AA\ channel (together with AIA 193\,\AA) simply for aligning XRT with Hi-C.

One problem concerning the XRT data during the Hi-C flight is that they do not fully cover the whole field-of-view of Hi-C. In particular, they only partially cover the plage region hosting the loop-like features. Still, the XRT image shown in Fig.\,\ref{fig_i}, which is taken during the Hi-C flight, fully covers the region between the negative magnetic polarity of the plage region and the positive polarity in the main part of the active region to the North. This is clear from the comparison to the HMI magnetogram in Fig.\,\ref{fig_i}. Therefore XRT covers the region where one would expect the hot loops that might be rooted in the plage region ($\beta$ in Fig.\,\ref{fig_i}).

In the space covering the connection from the plage area with the small loop-like features to the main part of the active region, there is no significant X-ray emission visible  (Fig.\,\ref{fig_i}). While we see clear hot loops in X-rays in the main part of the active region that contains the moss regions already studied by Testa et al. (2013), such X-ray emission is not related to the plage region ($\beta$ in Fig.\,\ref{fig_i}).
In fact, the region North of the plage region, where the magnetic connection would be expected, is particularly dark in X-rays. This extends the temperature range of the missing hot plasma related to the plage region with the small loop features and highlights the lack of significant amounts of plasma present at higher temperatures from above 2\,MK up to 10\,MK. 

With these considerations on the thermal structure based on the DEM and the X-ray emission, we conclude that it is unlikely that  the loop-like features are the footpoints of hot (and dense) loops. Therefore, we exclude the moss scenario shown in Fig.\,\ref{fig_0}b and discussed in the introduction.

\section{Discussion}\label{S:discussion}

The above discussion of the thermal structure suggests that the loop-like features are not the moss-like footpoints of hot dense loops. This argument is based on the absence of a significant amount of hot plasma that would fill the related loop. Still, one could speculate that the scenario in Fig.\,\ref{fig_0}c could apply; namely that the small loop-like features are at the footprints of larger (hot) but dilute loops (cf. Sect.\,\ref{S:introduction}). Heat input from below could heat the lowermost part, but leave the upper part of the loop unaffected. In this case, the longer loops would not be visible in X-rays or leave a signature in the DEM at high temperatures, while we still see the compact loop-like features appearing at their footpoints. In principle, this would be consistent with the thermal structure we discussed in Sect.\,\ref{S:thermal.structure}.

However, there are theoretical arguments against this interpretation of the small loop-like features being footpoints of long dilute loops. To illustrate this we conduct a Gedanken experiment:  We assume that along a magnetic fieldline with a length of several tens of  Mm close to the footpoint exists a dense structure at approximately  1.5 MK with a length of approximately  1\,Mm along the fieldline. Above that, the loop is hotter but very dilute%
\footnote{Alternatively, the upper part of the loop could also be cool, say at only 10$^4$\,K. Because the Barometric scale height at low temperatures (at 10$^4$\,K it would be only 500\,km) is small compared to the loop length, also in this case the loop would be at very low density in its upper part. Therefore, for a cool upper part of the loop the arguments would be the same as for a dilute hot upper part discussed here, and our conclusion would be identical.}
 This structure would be, in general, consistent with our observation (according to the scenario in Fig.\,\ref{fig_0}c). We would see an elongated feature with a length of 1\,Mm and with a peak of the DEM at approximately ${\log}T{\approx}6.2$ or 1.5\,MK (cf.\ Fig.\,\ref{fig_h}), but no X-ray emission above, because the hot part of the loop is not dense enough. However, the thermal stratification assumed for this setup would not be stable over the time of one to two minutes when we observe the features during the Hi-C flight. The dense plasma at a 
 temperature of ${\log}T{\approx}6.2$ 
or 1.5 MK has a pressure scale height of approximately  75\,Mm (or 0.1 solar radii). With no (significant amount of) dense plasma above it, the dense 1.5\,MK plasma would quickly expand into the upper part of the loop, similarly to an expansion into vacuum. Expansion into vacuum happens at approximately the adiabatic sound speed, which for plasma at 1.5\,MK is approximately  185\,km\,s$^{-1}$. Even when assuming an expansion speed much less than that, 100\,km\,s$^{-1}$, for example, within the time-scale during which these structures exist (and we observe them), that is, approximately  100s, they should have expanded by approximately  10\,Mm (${=}~100$\,km\,s$^{-1}$ $\times$ 100\,s). This would be a distance much larger than the size of the structures, an expansion we should clearly see (e.g. in the form of a jet). This contradicts our initial assumption of 1.5\,MK dense plasma below a hot dilute loop, because we do not see any significant expansion of the small loop-like features. This leads us to the conclusion that they cannot be the footpoint regions of large dilute loops.    

This leaves only one scenario, namely that the small loop-like features
are indeed tiny versions of hot coronal loops, as depicted in Fig.\,\ref{fig_0}a.
The main argument against this could be that in the HMI magnetograms we see only one single polarity in the plage region hosting these structures, as already pointed out by \citet{2013A&A...556A.104P}. However, as discussed in Sect.\,\ref{S:magnetic.field}, these HMI observations of only one single polarity do not rule out the possibility of small-scale opposite polarities, which is because of the limited sensitivity and spatial resolution; instruments with higher resolution and sensitivity have seen such small-scale opposite polarities (cf. Sect.\,\ref{S:introduction}), even though no such observations are available during the Hi-C flight for the area we investigate.

There is also observational evidence for small loops at coronal temperatures through data from AIA; albeit those loops are  significantly longer than the miniature loops we report here. \citet{2016ApJ...820L..13W}
finds  short coronal loops in AIA 171\,\AA\ observations in plage regions that have footpoint distances from 3\,Mm to 5\,Mm. This implies that their lengths are of the order of 5\,Mm to 10\,Mm, depending which loop geometry is assumed. These loops also occur in plage regions that appear unipolar in HMI, and \citet{2016ApJ...820L..13W}
uses this as indirect evidence for small-scale opposite polarity magnetic field hidden to HMI.
In
some way, the small loops in plages found recently by \citet{2016ApJ...820L..13W}
could be considered as longer versions of the structures 
we study here in some detail  and were first reported by \citet{2013A&A...556A.104P}. In terms of modelling, it would be interesting to see what one-dimensional loop models would predict for very short loops that are heated to 1.5\,MK or more, if that is possible. The models by \citet{2012A&A...537A.150S} produce transition region loops with temperatures well below 1\,MK only. If, in those models, the energy input is increased, the loops might reach higher temperatures, and it would be interesting to see if they turn out to be stable or highly intermittent.

\section{Conclusions}\label{S:conclusions}

We investigated small loop-like features in the solar corona seen in the 193\,\AA\ channel of Hi-C. They have typical lengths of 1\,Mm and lifetimes of some 100s. They appear in a plage or enhanced network region at the periphery of an active region.  The analysis of the morphology of these structures in terms of horizontal motions and lifetime suggests a close relationship with the granular motions in the photosphere (Sect.\,\ref{S:prop.detail}).
One end of the
loop-like features is rooted in a magnetic field concentration as seen by HMI. The limited resolution and sensitivity of HMI prevents definite conclusions if the other end of the feature is rooted in an opposite magnetic polarity (Sect.\,\ref{S:magnetic.field}).

The analysis of the thermal structure of the loop-like features  (Sect.\,\ref{S:thermal.structure})
shows that they cannot be the footpoints of hot dense loops, as is the case for moss (cf.\,Fig.\ref{fig_0}b). This is supported by a DEM analysis (Sect.\,\ref{S:DEM}) as well as by the direct X-ray observations and a comparison to actual moss structures (Sect.\,\ref{S:X.rays}).
Theoretical arguments can be made that the loop-like features are also not at the footpoints of (hot) dilute loops  (cf.\,Fig.\ref{fig_0}c)
heated
from their footpoints, because then we would expect to observe expanding (jet-like) features, which we do not see (Sect.\,\ref{S:discussion}).

This discussion leaves us interpreting the loop-like features as actual miniature coronal loops (cf.\,Fig.\ref{fig_0}a). These would span just a single granule and connect a magnetic concentration in an intergranular lane with a feature of opposite-polarity on the other side of the granule. That HMI does not see  a parasitic opposite polarity does not exclude this scenario, because high-resolution photospheric magnetic field observations \citep[e.g.][]{2009ApJ...700.1391M,2013SoPh..283..253W}, as well as indirect evidence from coronal structures \citep{2016ApJ...820L..13W},
provide evidence for the existence of such small-scale opposite polarities (Sects.\,\ref{S:magnetic.field} and \ref{S:discussion}).

These miniature loops might well be related to small-scale flux emergence events, where small magnetic fluxtubes break through the photosphere  \citep[e.g.][]{2010ApJ...713.1310I}. Clear evidence for such  short (1Mm) loops reaching transition region temperatures (i.e.\ approximately  0.1\,MK) has been reported recently using IRIS data \citep{2014Sci...346E.315H}. It does not seem unrealistic that, in cases of stronger heating, such loops might reach higher coronal temperatures of more than 1\,MK. The transition region loops are visible by IRIS because of its high spatial resolution of approximately  0.3\arcsec. To see such miniature loops in the corona, one needs comparable spatial resolution, which is not available with the current workhorse of coronal imaging, AIA, but is made available by the rocket experiment Hi-C that we used here.
For this (and for other reasons) we hope for a reflight of Hi-C.

\begin{acknowledgements}
We very much appreciate the valuable comments of the anonymous referee. We would like to thank Philippe Bourdin for advice on the fitting of ellipses. Thanks are also due to Leping Li and Chen Nai-Hwa for helpful suggestions.  
This work was supported by the International Max-Planck Research School (IMPRS) for Solar System Science at the University of G{\"o}ttingen.
\end{acknowledgements}

\bibliographystyle{aa}
\bibliography{ad}

\Online

\begin{appendix}

\section{Properties of individual loop-like features}\label{S:app.properties}

\begin{figure*}
   \centering
  \includegraphics[width=18cm]{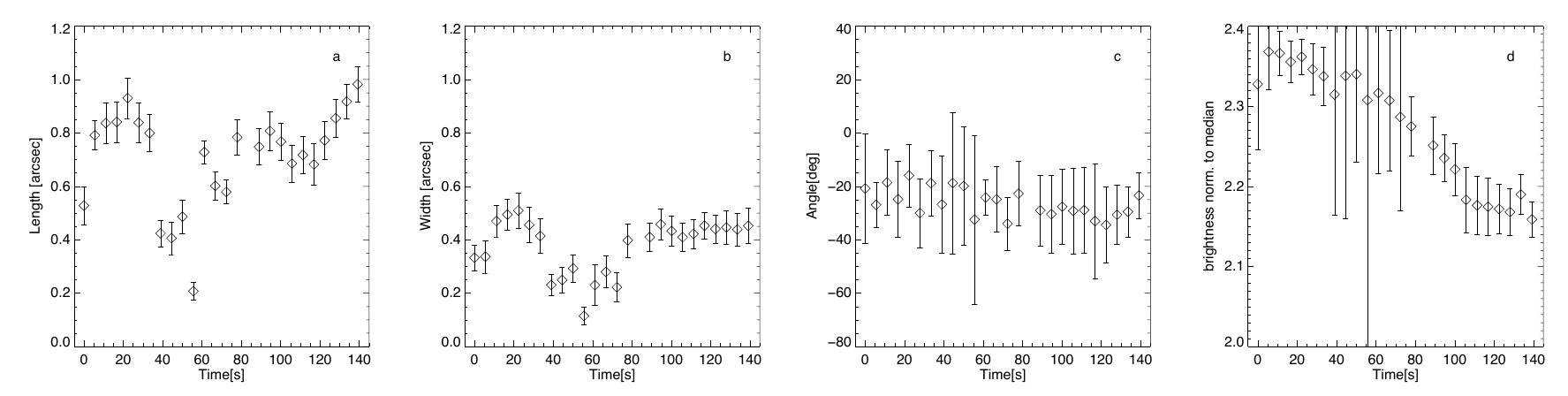}
      \caption{Properties of loop-like feature A marked in Fig.\,\ref{fig_b}. Same format as Fig.\,\ref{fig_e}}
         \label{str_a}
   \end{figure*}

\begin{figure*}
   \centering
  \includegraphics[width=18cm]{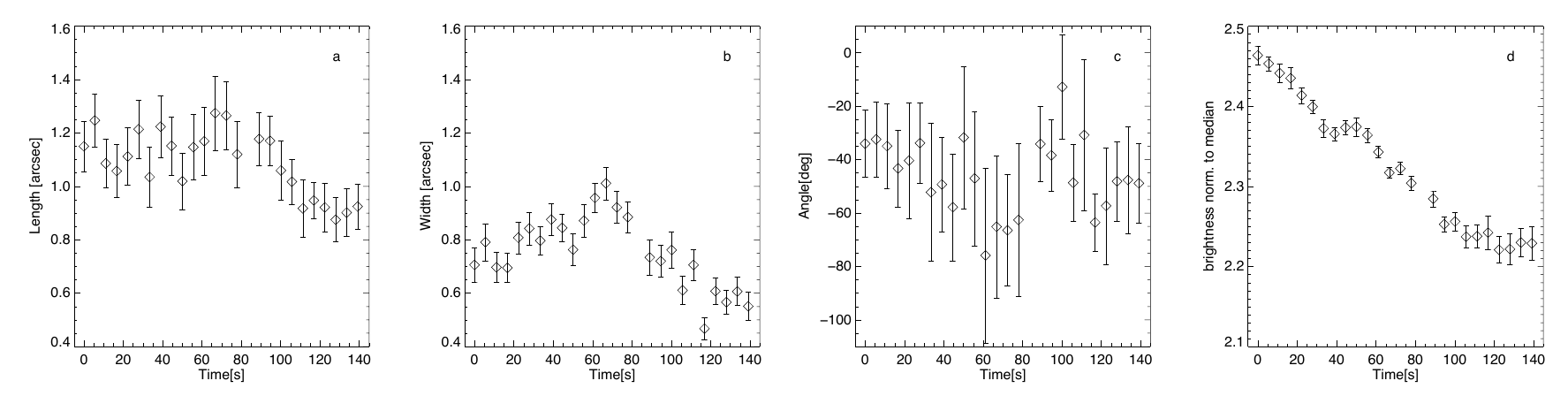}
      \caption{Properties of loop-like feature C marked in Fig.\,\ref{fig_b}.
Same format as Fig.\,\ref{fig_e}}
         \label{str_c}
   \end{figure*}

\begin{figure*}
   \centering
  \includegraphics[width=18cm]{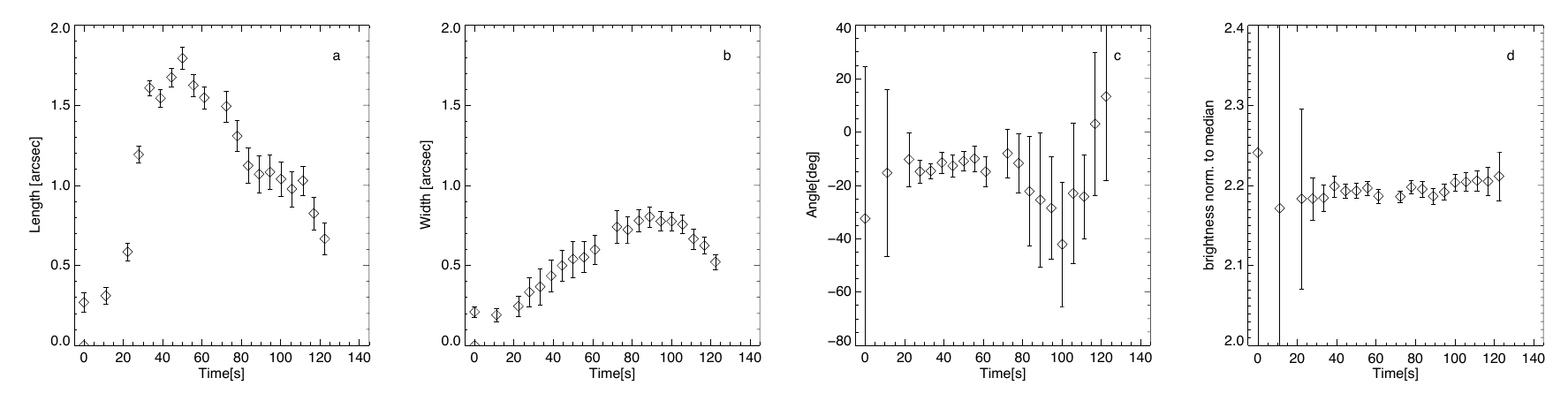}
      \caption{Properties of loop-like feature D marked in Fig.\,\ref{fig_b}.
Same format as Fig.\,\ref{fig_e}}
         \label{str_d}
   \end{figure*}

In Sect.\,\ref{S:prop.detail} we discussed the properties of the loop-like structures in terms of length, width, orientation and brightness. There we mainly concentrated on one single structure, namely feature B marked in Fig.\,\ref{fig_b}. Here we add the corresponding properties for the three additional structures, A, C and D, as marked in Fig.\,\ref{fig_b}. In the same format as used in Fig.\,\ref{fig_e} for the time evolution of the properties of feature B, we show here in Figs.\,\ref{str_a} to \ref{str_d} the properties of features A, C and D.

The properties of all four features are approximately similar, except for the evolution of the brightness. We could follow only feature B over a full life-cycle, that is, from an increase of the brightness until it faded away again within less than 2\,min (cf.\,Fig.\,\ref{fig_e}d). The three other features shown here are seen either only in the declining phase (Figs.\,\ref{str_a} and \ref{str_c}) or show an almost constant brightness (Fig.\,\ref{str_d}).

%-----------------------------------------------------------------------
   
%----------------------------------------------------------------------

\newpage

\section{Spatial alignment of Hi-C with magnetogram}\label{S:alignment}
 
To investigate the relation of the small loop-like coronal structures to the magnetic field in the photosphere we have to perform an alignment through different steps. 
The result of this alignment procedure is shown in Fig.\,\ref{fig_f} and discussed in Sect.\,\ref{S:magnetic.field}.
In this appendix we briefly describe the alignment procedure.

\begin{figure*}
   \centering
  \includegraphics[width=18cm]{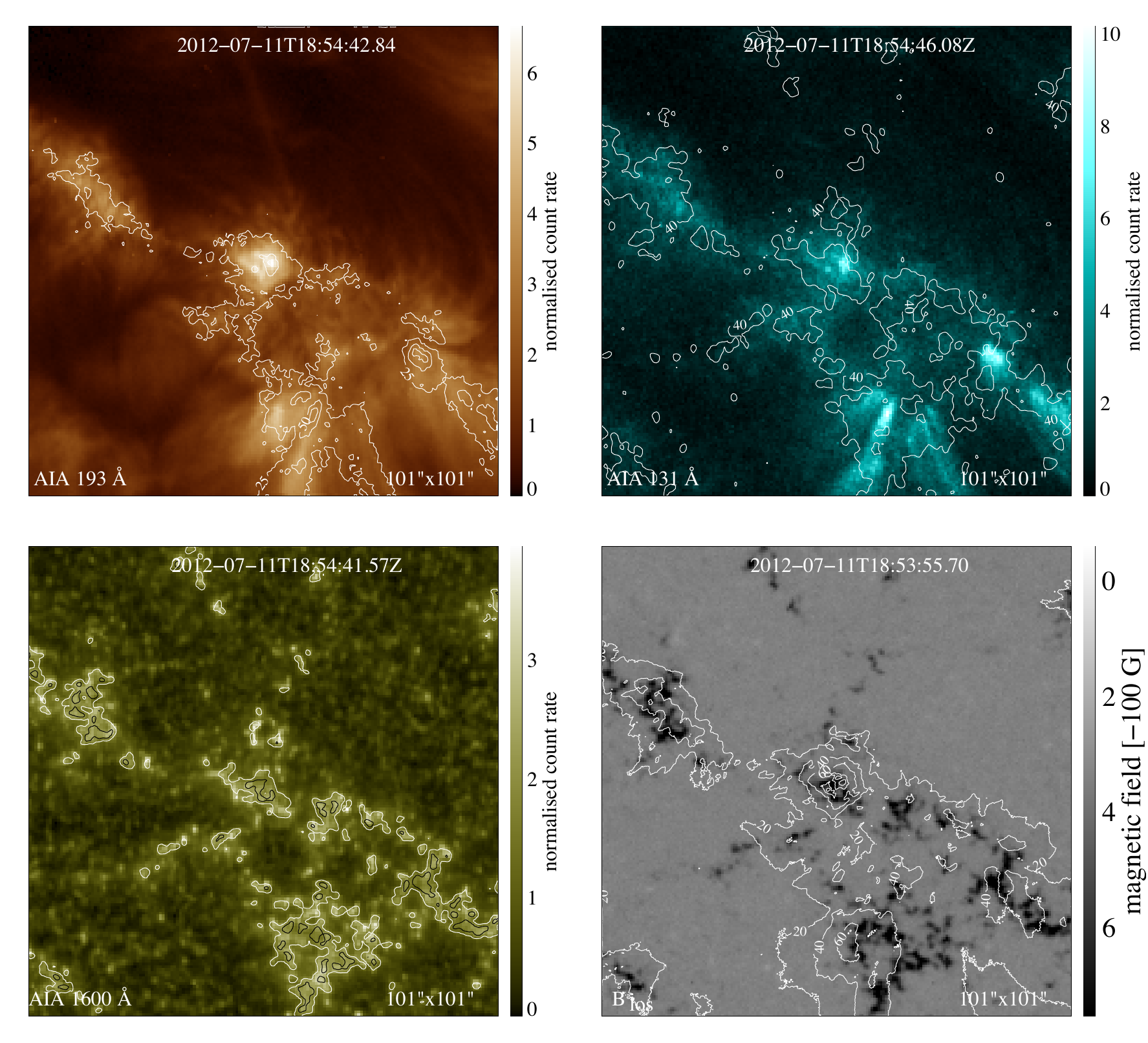}
      \caption{Illustration of the alignment procedure.
The panels show the sequence of 101\arcsec$\times$101\arcsec\ images of the alignment chain, that is, AIA\,193\,\AA\ $\to$ AIA\,131\,\AA\ $\to$ AIA\,1600\,\AA\ $\to$ HMI magnetogram.
In each image the contours of the respective image following in the chain are overplotted.
The Hi-C 193\,\AA\ image is aligned with the AIA\,193\,\AA\ channel and its contours are overplotted on the HMI magnetogram.
The magnetogram contours are at $-100$\,G (white) and $-300$\,G (black). The other contours show the percentages of the maximum brightness in the respective channel (numbers with the contour lines).
See Appendix\,\ref{S:alignment}
      }
         \label{app_aa}
   \end{figure*}

For the alignment chain we use a subregion of approximately  100\arcsec$\times$100\arcsec\ with the plage region hosting the small loop-like structures at approximately the center.
All images used for the alignment are scaled (through interpolation) to the pixel scale of Hi-C.
We first align the Hi-C image to the AIA image in the same 193\,\AA\ channel.
Because these images show the same coronal emission, only at different spatial resolution, this alignment is very reliable.
The next step is to align the AIA 193\,\AA\ image to the AIA 131\,\AA\ channel.
While the temperature of the peak contribution is quite different (cf.\ Table\,\ref{table:table1}), both channels also  have a significant contribution from lower (transition region) temperatures from ${\log}T{=}5.4$ to $5.6$ \citep[cf.][]{2012SoPh..275...41B}.
Therefore, the plage and enhanced network regions around our region-of-interest\ show many similar structures in the emission in the 193\,\AA\ and the 131\,\AA\ that provide a good basis for the alignment.
This is also the case for the next step when aligning the AIA 131\,\AA\ channel with the AIA 1600\,\AA\ images that show the low chromosphere, because in both channels the emission from the (enhanced) network will dominate.
In the final step we align the AIA 1600\,\AA\ image with the HMI magnetogram.
This alignment procedure is illustrated in Fig.\,\ref{app_aa}, where we show the respective images together with contour plots of the image in the next step of the alignment procedure.

For all alignment steps we use a cross-correlation technique to determine the misalignment and then shift the images with sub-pixel accuracy.
We test the accuracy of the alignment through an alignment of the final aligned images.
In Table\,\ref{table:2} we list the misalignment (separate for the solar $x$ and $y$ directions) of all alignment steps.
Combined, these give a misalignment of 0.2\arcsec\ and 0.3\arcsec\ in the $x$ and $y$ directions, respectively.
This alignment accuracy is of the order of the spatial resolution of Hi-C, or a fraction of a pixel of HMI or AIA, just as one would expect.
This emphasises the fact that we can trust the spatial alignment between Hi-C and the HMI magnetograms, which is the basis for the discussion on the relation of the small loop-like features to the magnetic field in Sect.\,\ref{S:magnetic.field}.

\begin{table}[h]
\caption{Accuracy of spatial alignment in solar $x$ and $y$ directions.}             % title of Table
\label{table:2}      % is used to refer this table in the text
\centering                          % used for centreing table
\begin{tabular}{r @{$~~\leftrightarrow$~~} l c c c}        % centred columns (4 columns)
\hline\hline                 % inserts double horizontal lines
\multicolumn{2}{c}{images} & $x$~[\arcsec] & $y$~[\arcsec] \\    % table heading 
\hline                        % inserts single horizontal line
   HMI      & AIA 1600 & 0.06 & 0.13 \\      % inserting body of the table
   AIA 1600 & AIA 131  & 0.10 & 0.24 \\
   AIA 131  & AIA 193  & 0.13 & 0.18 \\
   AIA 193  & Hi-C     & 0.01 & 0.01  \\
\hline                                   %inserts single line
\end{tabular}
\end{table}

\end{appendix}

\end{document}